\documentclass[runningheads]{llncs}

\usepackage[T1]{fontenc}
\usepackage{graphicx}
\usepackage{acronym}
\usepackage{amsmath}
\usepackage{tabularx}
\usepackage{booktabs}
\usepackage{acronym}
\usepackage{makecell}
\usepackage{float}
\usepackage{url}
\usepackage{xspace}
\usepackage{siunitx}
\newcommand{\Vadc}{\ensuremath{\hat{V}}\xspace}%
\newcommand{\Va}{\ensuremath{V}\xspace}%
\newcommand{\amp}{\ensuremath{G}\xspace}%
\DeclareMathOperator{\fom}{FoM}
\DeclareMathOperator{\rank}{RANK}
\DeclareMathOperator{\modrank}{E-RANK}

\begin{document}

\title{A Comparative Analysis of Lightweight Hash Functions Using AVR ATXMega128 and ChipWhisperer}

%Performance Analysis of Lightweight Hash Functions on an AVR Microcontroller using ChipWhisperer}

%\title{Software Benchmarking and Comparative Performance Analysis of NIST-Submitted and ISO-Standardized Lightweight Hash Functions on an AVR Microcontroller}

% Blinded
\author{\vspace{2cm}}

\author{Mohsin Khan\inst{1}\thanks{Corresponding author: Mohsin Khan, email: mohsin.khan@uit.no}
\and
Dag Johansen\inst{1}
\and
H{\aa}vard Dagenborg\inst{1}
}

% \author{Mohsin Khan\inst{1}\orcidID{0000-0003-1815-8642}\thanks{Corresponding author: Mohsin Khan, email: mohsin.khan@uit.no}
% \and
% Dag Johansen\inst{1}\orcidID{0000-0001-7067-6477}
% \and
% H{\aa}vard Dagenborg\inst{1}\orcidID{0000-0002-1637-7262}
% }

\titlerunning{A Comparative Analysis of Lightweight Hash Functions}

%\authorrunning{M. Khan et al.}

\institute{}
\institute{UiT: The Arctic University of Norway, Troms\o, Norway \\
\email{\{mohsin.khan, havard.dagenborg, dag.johansen\}@uit.no}}

\maketitle              % typeset the header of the contribution

\acrodef{IoT}{Internet of Thing}
%\acrodef{LWHF}{Lightweight Hash Function}
\acrodef{CpB}[CPB]{Cycles per Byte}
\acrodef{FoM}[FOM]{Figure of Merit}
\acrodef{GE}{Gate Equivalent}

\begin{abstract}

Lightweight hash functions have become important building blocks for security in embedded and IoT systems.
A plethora of algorithms have been proposed and standardized, providing a wide range of performance trade-off options for developers to choose from.
This paper presents a comparative analysis of 22 key software-based lightweight hash functions, including the finalist from the SHA-3 competition.
We use a novel benchmark methodology that combines an AVR ATXMega128 microcontroller with the ChipWhisperer cryptanalysis platform and
evaluate and compare the various hash functions along several dimensions, including execution speed, % measured in Cycles per Byte (CpB), 
memory footprint, and energy consumption.   
Using the composite $\modrank$ metric, we provide new insight into the various trade-offs each hash function offers to system developers.

%Aside from the status reports provided by NIST, 
%However, few studies benchmark and compare their performance on a common platform.
%balance of resource consumption, computational efficiency, and energy consumption.
%evaluated through RAM and ROM consumption,
% We utilize $\modrank$ to provide 
%a balanced composite trade-off metric that offers insights into lightweight hash functions achieving 
%an optimal balance of resource consumption, computational efficiency, and energy consumption.
%We provide we conduct a detailed software benchmarking and comparative analysis of key lightweight hash functions 
%in comparison with ISO-standardized lightweight hash functions, including PHOTON and Lesamnta-LW, as well as the SHA competition finalist BLAKE.
%
%We perform the benchmarking on the AVR ATXMega128D4 microcontroller using the ChipWhisperer platform, 
%which is a comprehensive suite of hardware and software tools designed to analyze cryptographic primitives. 

\end{abstract}
\section{Introduction}

%They provide authentication through digital signatures and message authentication codes, data integrity and verification via checksums, 
%and secure networking and communication through protocols like SSL/TLS and the use of digital certificates. 
%%Hash functions are basic components of blockchain and help maintain data integrity and secure transactions. 

Hash functions are vital building blocks for many security mechanisms, like digital signatures, message authentication codes, and file checksums. 
Traditional hash functions are effective on devices with ample computational resources, but may
fall short on resource-constrained devices with limited computational capacity, such as 8-bit AVR and PIC family of microcontrollers.
In some computational environments, like IoT and embedded systems, devices might also be battery-powered and require energy-efficient algorithms.

%
%or that are powered by batteries, such as in IoT applications.
%In addition to applications like home automation and consumer electronics, 
%these devices also serve critical functions in sectors such as medical technology, automotive systems, and defense applications. 
%and memory, but resource-constrained devices, like IoT devices, sensors, 
%and RFIDs, face challenges such as high power consumption, significant memory usage, and increased latency. 
%There is a wide variety of resource-constrained devices that have limited computational power, limited memory availability, and limited power sources, 
%such as the ATmega, ESP32, and PIC16F84A microcontrollers. 
%Most of these microcontrollers operate in a bare-metal environment or utilize a very small operating system. 
%However, traditional hash functions are resource-intensive, making them impractical for use on resource-constrained devices. 
%The need for secure hash functions in these critical applications has led to initiatives to develop \acp{LWHF} that provide sufficient security while accommodating the limitations of these devices. 

%These algorithms are typically not developed to defend against side-channel attacks. 
%In contrast, \acp{LWHF} are optimized for the limitations of these devices, supporting various security applications efficiently. 

DM-PRESENT~\cite{dm_present}, to our knowledge, was among the first hash functions developed specifically for devices with limited resources.
From 2007 to 2012, NIST ran the SHA-3 competition to find a more lightweight and secure alternative to SHA-2. 
Keccak~\cite{Keccak} won this competition, with BLAKE as one of the finalists. 
Both of these hash functions are actively used in cryptographic libraries today, such as OpenSSL and Python's hashlib. 
Additionally, Keccak is specifically utilized in blockchain technology and smart contracts. 
In 2016, ISO standardized PHOTON, SPONGENT~\cite{SPONGENT}, and Lesamnta-LW under ISO/IEC 29192-5:2016. 
Then, in 2019, NIST began the standardization of lightweight cryptographic algorithms and received 57 submissions, some of which included lightweight hash functions.  
After three rounds of evaluation and rigorous analysis of security and benchmarking results, ASCON~\cite{dobraunig2021ascon} was announced as the winner in 2023. 
Consequently, a plethora of hashing functions have been proposed and developed, providing a wide range of performance and security trade-offs.
However, few works exist that analyze and compare a larger selection of these hash functions. 

This paper benchmarks and compares 22 key lightweight hash functions, including those submitted to NIST~\cite{NIST_LWC},
those standardized by ISO~\cite{ISO29192-5:2016},  
including PHOTON and Lesamnta-LW, along with the SHA-3 competition~\cite{chang2012third} finalist, BLAKE~\cite{blake2} 
and its latest variant, BLAKE3~\cite{BLAKE3}, due to its wide usability in various well-known applications. 

%(specifically BLAKE2s~\cite{blake2}  and BLAKE3). 
%We have selected 20 lightweight hash functions from the NIST submission, as well as PHOTON and Lesamnta-LW from ISO standardization, 
%as listed in Table~\ref{tab:LWHF_comparison}. 
%We are also considering the BLAKE2s variant from the BLAKE2~\cite{blake2} lightweight hash function, which was part of the SHA Competition,
%and its latest variant, BLAKE3~\cite{BLAKE3}, due to its wide usability in various well-known applications. 

For each hash function, we measure several performance metrics, including \ac{CpB}, energy consumption, RAM usage, and ROM using 
a novel benchmark environment comprised of the AVR ATmega128 microcontroller in combination with the ChipWhisperer~\cite{newae2025Documentation} 
cryptanalysis tool. 
We also calculate the composite $\modrank$ metric~\cite{10.1007/978-981-96-2417-1_14} to compare provide 
a more comprehensive overview of the performance trade-offs of each hash function.

%There is still a need for advanced research into software performance benchmarking, the trade-offs involved, and the evaluation of lightweight hash functions,
% particularly those submitted to NIST. 
%the software performance of lightweight hash functions, particularly those submitted to NIST and standardized by ISO,
%such as PHOTON~\cite{PHOTON} and Lesamnta-LW~\cite{Lesamnta_LW}.

%The hash functions are also compared based on how 
%overall performance based on how they balance resource consumption and energy efficiency using the 
%$\modrank$ metrics.

%These metrics are obtained using custom profiling modules designed for energy, memory, and execution time analysis. 
%Further, comparing these hash functions with the widely used BLAKE~\cite{BLAKE_book} hash, which was submitted 
%to the SHA-3 competition in 2008~\cite{chang2012third}, provides valuable insights into the effectiveness and efficiency of these hashing methods.

%\section{Background}

\section{ChipWhisperer}

\begin{figure}[tb]
    \centering
    \includegraphics[width=\linewidth]{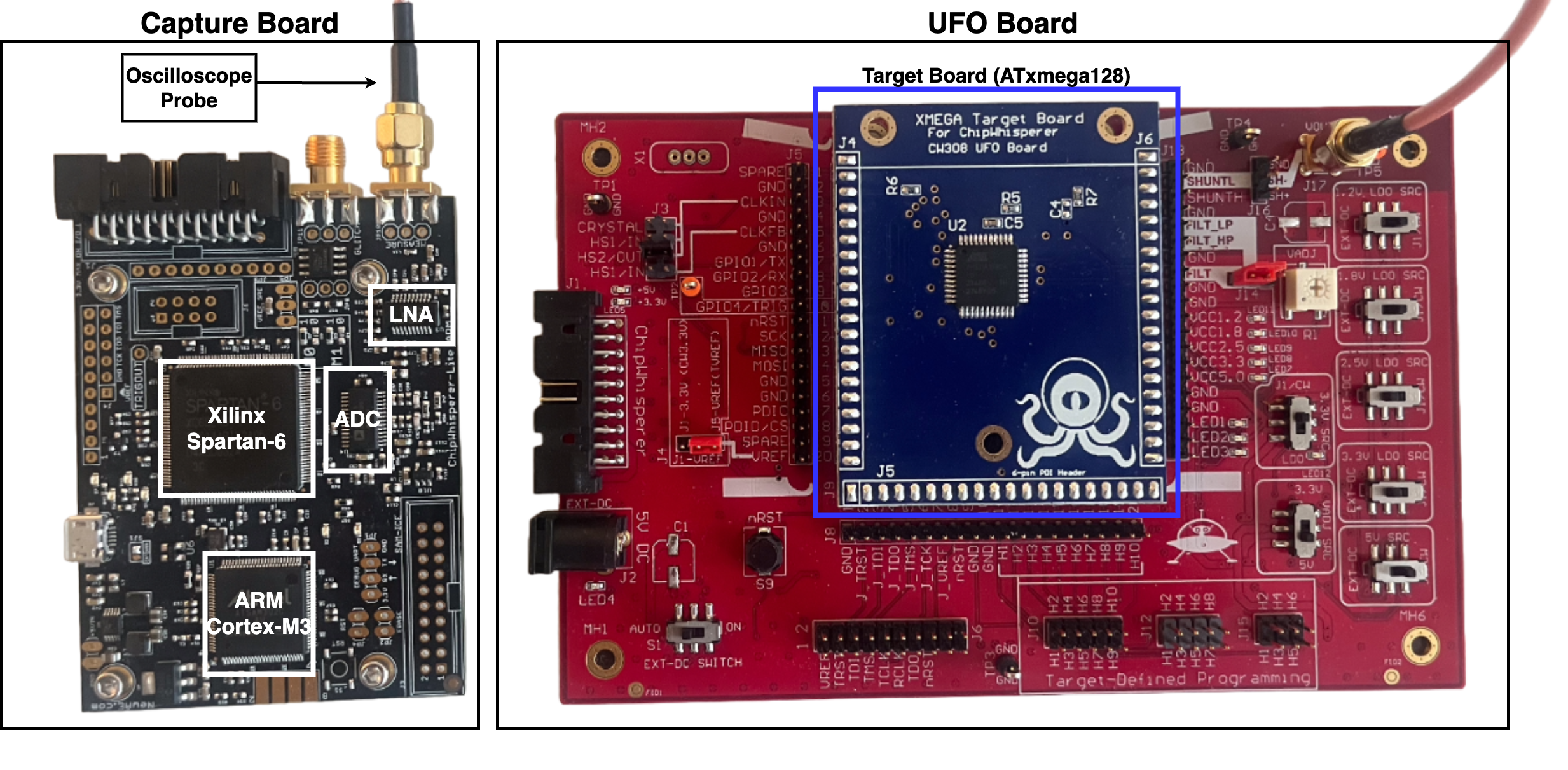} % Reduce size
    \caption{ChipWhisperer hardware overview}
    \label{fig:chipwhisperer_arch}
\end{figure}

Our measurements were performed using the ChipWhisperer Level 2 Starter Kit from NewAE Technology, a toolkit 
originally designed for learning about side-channel attacks on embedded devices by analyzing power consumption.
Its modular design allows for the integration of specialized modules, 
enabling precise measurements and advanced testing of cryptographic systems. 
ChipWhisperer comprises a \emph{capture board} and a \emph{target board}, 
connected via a serial connection, and an oscilloscope probe, as shown in Figure~\ref{fig:chipwhisperer_arch}.

The ChipWhisperer capture board includes an FPGA microcontroller, an Analog-to-Digital Converter (ADC), and a Low-Noise Amplifier (LNA). 
The FPGA microcontroller serves as the central processing unit, managing communication, timing, and data acquisition processes. 
The LNA amplifies weak power signals to reduce noise and enhance the accuracy of the power analysis obtained from the target board. 
These amplified signals are then captured by the ADC for conversion into digital signals. 
The target board is mounted on a Universal Feature Observation (UFO) Board, 
which provides a standardized interface for power, clock, and data connections to the target board.
Data is transferred between ChipWhisperer and the target board over a serial port, 
while the oscilloscope probe collects voltage traces as the target board executes specific processes triggered by this communication.

%\subsection{Hardware and Software Setup}

The ChipWhisperer Python API allows interaction with the ChipWhisperer FPGA, enabling programming of the target board, 
configuring clock settings, and triggering the target board to start and stop cryptographic operations. 
It also facilitates the capturing and transferring of power traces. 
Also, the Python API enables the configuration of ADC and LNA parameters, such as sampling rate and trigger settings, 
as well as adjustments to gain for capturing power traces and amplifying weak power signals.

\begin{figure}[tb]
    \centering
    \includegraphics[width=\textwidth]{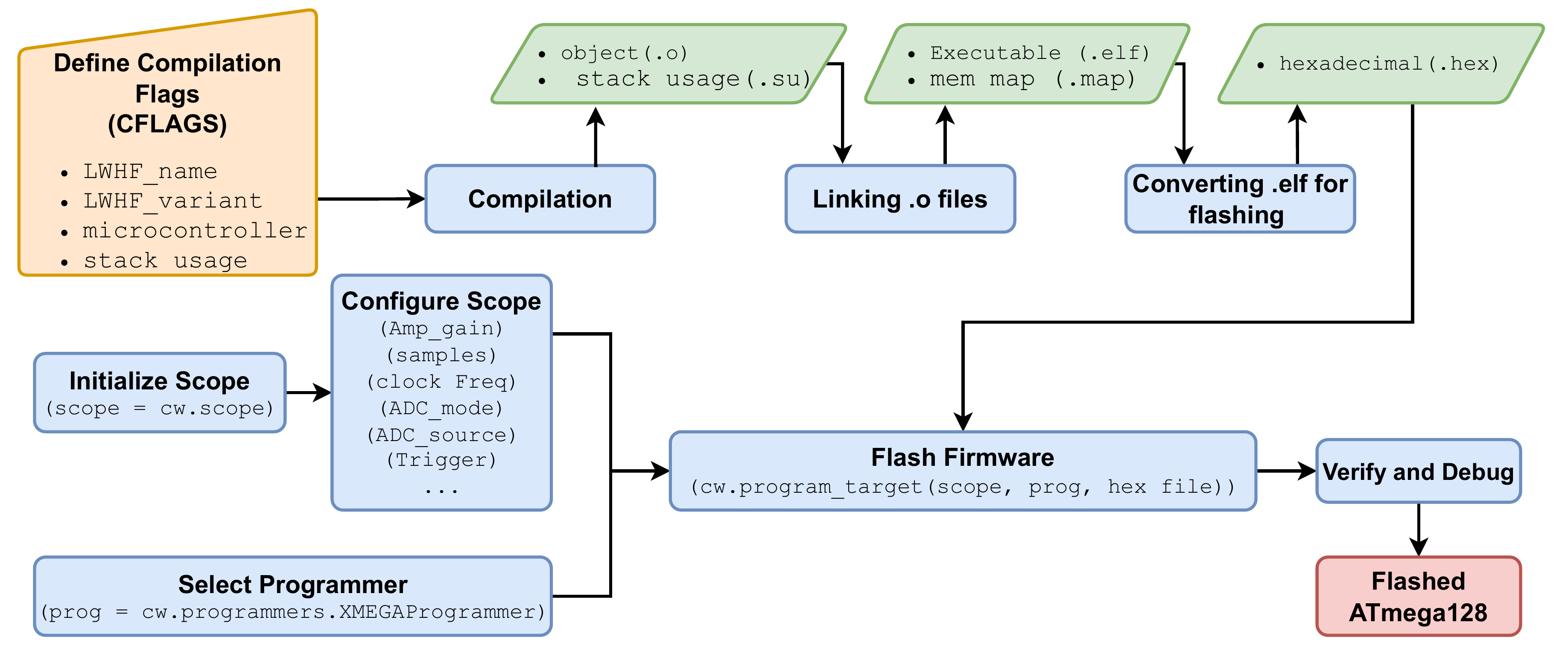}
    \caption{Process flow for firmware build and deployment to target device.} \label{Fig:firmware_compilation}
\end{figure}

The target board is programmed by uploading firmware, which is developed in the C language. 
This firmware includes a base C program designed to manage simple serial communication with the ChipWhisperer Python API. 
The base program handles trigger signals from the Python API and incorporates macros. 
These macros are used for the hash function and its variants, with the implementations of these hash functions sourced from their original publications. 
The firmware is compiled using AVR-GCC, including specific compiler flags for object files and macros that define the hash function name and variant. 
During the compilation process, the C implementations of the selected hash functions are converted into object files, and a corresponding \texttt{.su} file is generated to provide stack usage information. 
These object files are then linked together to create \texttt{.elf} and \texttt{.map} files. 
The \texttt{.elf} file is subsequently converted into a \texttt{.hex} file, which is used to flash the target device. 
The flashing is performed using the ChipWhisperer Python API, where the scope is initialized, and settings for amplifier gain, sampling rate, trigger conditions, and clock sources are configured. 
After flashing, the firmware is verified and debugged. The process for building firmware and deploying it on the target device using ChipWhisperer is illustrated in Figure~\ref{Fig:firmware_compilation}.

\begin{figure}[tb]
    \centering
    \includegraphics[width=0.8\textwidth]{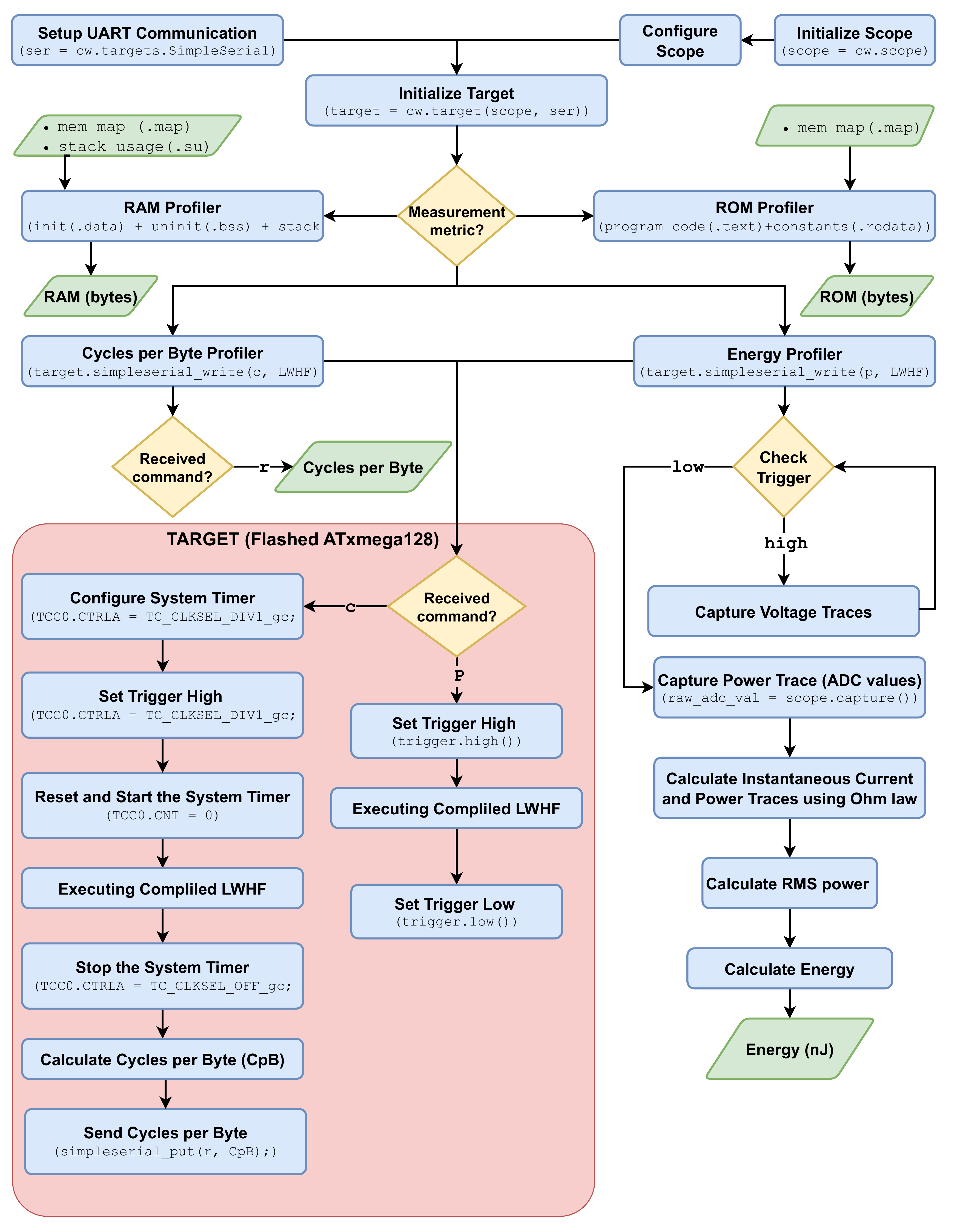}
    \caption{Benchmarking framework.} \label{Fig:benchmarking_framework}
\end{figure}

\section{Benchmarking Framework}

We have opted for the AVR ATXMega128D4 microcontroller~\cite{AtmelXMEGA_D4} 
as our target board for benchmarking the selected lightweight hash functions. 
ATXMega128D4 is a low-power 8-bit RISC microcontroller %with a Harvard architecture, 
with
 128\,KB of flash memory, 8\,KB of SRAM, 2\,KB of EEPROM, a clock frequency of 32\,MHz, and an operating voltage range of 1.6 to 3.6\,V. 
These resource constraint specifications are sufficient to implement all the lightweight hash functions in this study 
and conduct performance tests. 
This setup also allows precise control over code execution and simplifies optimization at the assembly level. 

Benchmarking begins by initializing the scope and configuring its settings. 
A 5\,dB gain is applied to amplify low-amplitude power signals to minimize distortion, allowing clear capture of signal variations during hash function operations. 
A high-to-low trigger edge synchronizes the capture window with the start and the end of the hash function execution, 
and a sampling rate of 5 samples during the capture window offers moderate resolution for accurate trace alignment. 
The clock frequency is set to 7.3728\,MHz using the ATXMega128D4's internal clock. 
Simultaneously, a simple serial communication is established using UART. 
After setting up the simple serial interface and the scope, the target device is initialized. 
Once the target initialization is complete, specific measurement metrics are selected 
for execution time, memory footprint, and energy profiling, 
as illustrated in the workflow of the benchmarking framework in Figure~\ref{Fig:benchmarking_framework}.

\subsection{Execution Time Profiling}

We measure execution time in terms of the number of processing cycles required to hash each byte of data, commonly known as \acf{CpB}.
Our \ac{CpB} profiler utilizes a control register associated with the timer/counter on the ATXMega128D4 target board. 
According to the hardware specifications for the ATXMega128D4 microcontroller,
the \texttt{TCC0.CTRLA} register is responsible for configuring the clock source and pre-scaler settings. 
When \ac{CpB} measurement is selected, 
the \texttt{TCC0.CTRLA} register is initialized and set to zero at the beginning of the hashing process. 

Once the hashing process has completed processing the input data, 
the control register is stopped and the total number of cycles used is retrieved from the register. 
The resulting value is divided by the total number of input bytes processed to compute the final \ac{CpB} value, before being transmitted to the host system.
Note that the pre-scaler is set to \texttt{CLK\_DIV1}, meaning that the timer/counter operates at the system clock frequency without any division.

\subsection{Memory Footprint Profiling}

In sponge constructions~\cite{Bertoni2007Sponge}, the internal state is divided into two parts: rate and capacity. 
The rate indicates the number of bits of the state that can be directly read or written during the absorption and squeezing phases, determining the throughput of the function. 
The capacity is the portion of the state that remains hidden from input and output, providing security against cryptanalytic attacks. 
Merkle–Damg{\aa}rd constructions~\cite{damgaard1989design}~\cite{merkle1989certified} do not explicitly separate the state into rate and capacity. 
Instead, the internal state is represented by the chaining value, while the rate is equal to the message block size. 
The overall state is updated by a compression function, 
and security is ensured through the design of that function and the padding scheme, rather than through a reserved hidden portion of the state.

For our purpose, we assess the memory footprint of each hash function by measuring its RAM and ROM consumption, which 
involves evaluating the size of the various memory segments used and summing up the total byte count.
For RAM, we include the sizes of both initialized and uninitialized global and static variables, along with both dynamic and static stack usage, as shown below.
\begin{equation}
    \label{eq:ram}
    \text{RAM} = \texttt{.data} + \texttt{.bss} + \texttt{.su} 
\end{equation}
The initialized \texttt{.data} segment and the uninitialized \texttt{.bss} segment are extracted from the \texttt{.map} file that was generated by the AVR-GCC compiler during the linking stage. 
Both dynamic and static stack usage data are obtained from the \texttt{.su} files created during the compilation of individual object files, providing a comprehensive breakdown of memory allocation. 
Heap memory allocated via \texttt{malloc} and \texttt{calloc} calls are excluded because dynamic memory allocation is not utilized in the C implementations of the selected hash functions. 

For ROM, we sum the memory occupied by the program code \texttt{.text} segment and constant data in the \texttt{.rodata} segment, as shown below. 
\begin{equation}
    \label{eq:rom}
    \text{ROM} = \texttt{.text} + \texttt{.rodata}
\end{equation}
The generated \texttt{.map} file provides a detailed memory map of the program code and constant data, allowing precise measurement of the total ROM footprint.

\subsection{Energy Profiling}

To profile energy consumption, we use the \texttt{cw.capture\_trace} API function of the ChipWhisperer toolkit. 
The API call returns an array of instantaneous voltage samples collected by the ADC, normalized to the range -0.5 to +0.5. 
The voltage samples are measured when a high-to-low trigger signal is detected from the target device via serial communication. 

While normalization ensures consistency across different hardware configurations, it requires us to convert the samples back to actual voltages. 
For this, let  
\Vadc be the normalized ADC voltage sample, \( V_{\text{ref}} \) the reference voltage of the ADC, which is in our case is \SI{1}{\volt}, 
 and \amp the gain of the amplifier applied to the signal before digitization, 
 which in our case is \SI{5}{\decibel}, then the  actual voltage \Va is given by 
%
%To convert the normalized values back to the actual voltage trace, we use Equation~\eqref{eq:vol_trace}. 
%In this equation, \( V_{\text{actual}} \) represents the actual voltage trace,
%
\begin{equation}
\label{eq:vol_trace}
\Va =  \frac{V_{\text{ref}}}{\amp} \times \Vadc =  \num{0.2}  \Vadc
\end{equation}

%\begin{equation}
%    \label{eq:vol_trace}
%    V = s \times \frac{1}{G} + V_{\text{ref}} =  \frac{1}{5} s + V_{\text{ref}}
%\end{equation}

%
Furthermore, for the ATXMega128D4 microcontroller, 
we have the shunt resistance  $R_{\text{shunt}} = \SI{49.9}{\ohm}$ and the supply voltage \( V_{\text{sup}} = \SI{3.3}{\volt}\).
The instantaneous current \( I \) of a sample \Vadc is then given by Ohm's law as follows
\begin{equation}
    \label{eq:curr_trace}
    I = \frac{\Va}{R_{\text{shunt}}} = \frac{\Va}{49.9} = \frac{0.2 \Vadc}{49.9} =  4.0 \times 10^{-3} \Vadc  %0.004008016032064129
\end{equation}
This gives us the instantaneous power \( P \) for sample \Vadc as follows.
\begin{equation}
    \label{eq:power_trace}
    P = I \times V_{\text{sup}} =   
    4.0 \times 10^{-3} \Vadc   \times 3.3 = 
    1.3 \times 10^{-2} \Vadc %    0.0132264 \Vadc
\end{equation}

%\( V_{\text{sup}} \) for the ATXMega128D4 microcontroller is 3.3V.

%The instantaneous current is measured using Ohm's law, as shown in Equation~\ref{eq:curr_trace}. 
%This instantaneous current is then used to calculate the power trace, as detailed in Equation~\ref{eq:power_trace}. 
%In these equations, \( R_{\text{shunt}} \) represents the value of the shunt resistor, and \( V_{\text{sup}} \) is the supply voltage. 
%The \( R_{\text{shunt}} \) is 49.9~$\Omega$, and the \( V_{\text{sup}} \) for the ATXMega128D4 microcontroller is 3.3V.

The values in the trace array can sometimes be negative due to signal oscillations present in the original voltage trace. 
To obtain accurate power and energy calculations, we therefore calculate the Root Mean Square (RMS) of the trace. 
Given a trace $[\Vadc_1, \Vadc_2, \ldots, \Vadc_N]$ of $N$ samples, 
the RMS power \( P_{\text{rms}} \) is given as follows.

%  0.0132264**2 = 0.00017493765695999997 = 1.75e-04 = 1.75\times10^{-4}
\begin{equation}
\label{eq:rms_power}
P_{\text{rms}} = \sqrt{\frac{1}{N} \sum_{i=1}^{N} P_{i}^2} = \sqrt{\frac{1.7\times10^{-4}}{N} \sum_{i=1}^{N}  \Vadc_i^2} 
\end{equation}

To calculate an accurate energy estimation, execution time needs to be considered, 
which represents the total duration the microcontroller spends executing the hash function. 
The execution time is determined using Equation~\ref{eq:execution_period}, where \( T_{\text{exec}} \) denotes the 
execution period, \( C \) is the total number of processing cycles used during the hash function's execution, and \( f_{\text{clk}} \) 
indicates the microcontroller's clock frequency, which in our case is set to \SI{7.3728}{\mega\hertz}. 
Once the execution period has been measured, the energy consumption can be calculated using Equation~\ref{eq:energy}.

\begin{equation}
\label{eq:execution_period}
T_{\text{exec}} = \frac{C}{f_{\text{clk}}} = \frac{C}{\SI{7.3728}{\mega\hertz} }
\end{equation}

\begin{equation}
\label{eq:energy}
E =   T_{\text{exec}} \times P_{\text{rms}} =  \frac{C}{\SI{7.3728}{\mega\hertz}} \times \sqrt{\frac{1.7\times10^{-4}}{N} \sum_{i=1}^{N}  \Vadc_i^2} 
\end{equation}

% \begin{equation}
% \label{eq:energy}
% Energy = \sqrt{\frac{1}{N} \sum_{i=1}^{N} P_{\text{trace}, i}^2} \times \frac{C}{f_{\text{clk}}}
% \end{equation}

\subsection{Performance Comparison and Ranking}

While single-dimensional metrics, such as throughput, \ac{CpB}, and energy consumption, are useful when comparing hash functions with similar optimization goals, 
they are not suitable for evaluating tradeoff between various performance dimensions.
For hardware implementations, \ac{FoM}~\cite{badel2010armadillo} is commonly used as compound metric
that measures performance as the ratio of throughput to the square number of logical gates (i.e., \acs{GE}).
\ac{FoM} also captures energy requirements, as power consumption is proportional to the number of logic gates, represented by the \ac{GE} factor.
Later variants of \ac{FoM} also normalize to the hardware clock frequency~\cite{10.1007/978-3-642-23951-9_22}, as shown in Equation~\ref{eq:fom}.
\begin{equation}
    \label{eq:fom}
    \fom = \frac{\text{throughput}}{\text{clk} \times \text{GE}^2}
\end{equation}

In our case, for software implementations, the $\rank$ metric~\cite{10.1007/978-3-319-16363-5_1} % in Equation~\ref{eq:RANK} 
is commonly used to measure the tradeoff between execution efficiency, CPB, relative to memory usage (i.e., RAM and ROM).
%
%\begin{equation}
%\label{eq:RANK}
%\rank = \frac{ 10^9 / C_{\text{pB}} }{(\text{ROM} + 2 \times \text{RAM})}
%\end{equation}
%
However, unlike \ac{FoM}, the included memory footprint of $\rank$ does not offer any insights into power dissipation. 
To mitigate this limitation, we instead use the more recent $\modrank$ metric~\cite{10.1007/978-981-96-2417-1_14}, 
as defined in Equation~\ref{eq:mod_RANK},  
which incorporates energy consumption $E$ for more detailed and accurate metric.
    
\begin{equation}
\label{eq:mod_RANK}
\modrank = \frac{ 10^9 / {\text{cpb}} }{(\text{ROM} + 2 \times \text{RAM}) \times E}
\end{equation}

\section{Results and Analysis}

\begin{table}
    \centering
    \scriptsize
    \caption{Overview of selected lightweight hash function}
    \label{tab:LWHF_comparison}
    \resizebox{\textwidth}{!}{%
        \begin{tabular}{l c c c c c c}
    \toprule
    \textbf{Name} &
    \makecell{\textbf{Rate} \\ \textbf{(bits)}} &
    \makecell{\textbf{Capacity} \\ \textbf{(bits)}} &
    \makecell{\textbf{State} \\ \textbf{(bits)}} &
    \textbf{Structure} &
    \makecell{\textbf{Internal} \\ \textbf{Primitive}} &
    \textbf{Rounds} \\
    \toprule
    PHOTON-256~\cite{PHOTON} & 32 & 256 & 288 & \makecell{Extended \\ Sponge} & \makecell{AES-like \\ permutation} & 12 \\
    \midrule
    Lesamnta-LW~\cite{Lesamnta_LW} & 128 & — & 256 & \makecell{Merkle–Damgård \\ (MD)} & \makecell{AES-based block cipher \\ (LW1 mode)} & 64 \\
    \midrule
    BLAKE2s~\cite{blake2} & 512 & — & 256 & \makecell{HAIFA \\ (MD variant)} & ChaCha-inspired G function & 10 \\
    \midrule
    BLAKE3~\cite{BLAKE3} & 512 & — & 256 & Binary Tree & BLAKE2s compression function & 7 \\
    \midrule
    ASCON~\cite{dobraunig2021ascon} & 64 & 256 & 320 & Sponge & Bit-sliced permutation & 12/8 \\
    \midrule
    PHOTON-Beetle~\cite{bao2021photon} & 32 & 224 & 256 & Sponge & PHOTON-256 permutation & 12 \\
    \midrule
    Xoodyak~\cite{daemen2020xoodyak} & 128 & 256 & 320 & \makecell{Duplex \\ (Cyclist mode)} & \makecell{3×32-bit slices, \\ XOR/rotate/shift} & 12 \\
    \midrule
    KNOT~\cite{KNOT2019} & 32 & 224 & 256 & \makecell{Sponge/\\Duplex} & \makecell{SPN-style substitution \\ and diffusion} & 68 \\
    \midrule
    ORANGISH~\cite{ORANGE2019} & 128 & 128 & 256 & \makecell{Sponge \\ (JH mode)} & PHOTON256 permutation & 12 \\
    \midrule
    SHAMAS~\cite{Shamash2019} & 64 & 256 & 320 & \makecell{Sponge/\\Duplex} & \makecell{Bit-sliced permutation, \\ linear matrix mixing, \\ byte-wise rotations} & 12 \\
    \midrule
    SIV-Rijndael~\cite{SIVRijndael2562019} & 32 & 224 & 256 & \makecell{Modified \\ Sponge} & Rijndael256 permutation & 14 \\
    \midrule
    SIV-TEM-PHOTON~\cite{SIVTEMPHOTON2019} & 32 & 224 & 256 & \makecell{Modified \\ Sponge} & PHOTON-256 permutation & 20 \\
    \midrule
    SKINNY-tk2~\cite{SKINNY2019} & 32 & 224 & 256 & Sponge & SKINNY-128-256 TK Cipher & 48 \\
    \midrule
    SNEIKHA~\cite{SNEIK2019} & 256 & 256 & 512 & \makecell{Sponge \\ (BLNK2)} & \makecell{SNEIK f512 ARX \\ Permutation} & 8 \\
    \midrule
    TRIAD~\cite{TRIAD2019} & 32 & 224 & 256 & \makecell{Extended \\ Sponge} & Triad-P permutation & 1024 \\
    \midrule
    Coral~\cite{YararaCoral2019} & 32 & 224 & 256 & Sponge & $\pi\mathsf{I}$ permutation & 10 \\
    \midrule
    Gimli~\cite{Gimli2019} & 128 & 256 & 384 & Sponge & Gimli permutation & 24 \\
    \midrule
    CLX~\cite{CLX2019} & 32 & 256 & 288 & Sponge & $P'_{288,n}$ NLFSR permutation & var \\
    \midrule
    ACE-H~\cite{ACE2019} & 64 & 256 & 320 & \makecell{Sponge \\ (sLiSCP-light)} & \makecell{ACE Permutation \\ (Simeck-style)} & 48 \\
    \midrule
    ESCH~\cite{SPARKLE2019} & 128 & 256 & 384 & \makecell{Modified \\ Sponge} & ARX-based Sparkle384 & var \\
    \midrule
    Subterranean~\cite{Subterranean2019} & \makecell{32 (out) \\ 9 (in)} & 224 & 257 & \makecell{Flat Sponge \\ (Duplex)} & Bitwise round function & 1 \\
    \midrule
    Saturnin~\cite{Saturnin2019} & 256 & — & 256 & MD construction & Saturnin Block Cipher & 32 \\
    \midrule
    ISAP~\cite{ISAP2019} & 144 & 256 & \makecell{320 / \\ 400} & Sponge & Keccak-p[400] and Ascon-p & var \\
    \midrule
    GAGE~\cite{GAGE2019} & 8 & 224 & 232 & Sponge & Custom SPN permutation & 32 \\
    \bottomrule
    \end{tabular}%

}
\end{table}

In this section, we present the results of our benchmarking and performance analysis of the selected hash functions.
The list of all benchmarked hash functions is provided in Table~\ref{tab:LWHF_comparison}, including 
an overview of their internal parameters and structure.
State refers to the total amount of internal memory allocated for use during the hash function's operation.

The experimental results are visualized using graphs arranged in ascending order for execution time (\ac{CpB}), 
memory footprint (both RAM and ROM), and energy consumption, while the composite $\modrank$ metric is presented in descending order. 
This arrangement places the most efficient hash function  at the top, allowing for quick identification of optimal candidates. 
Performance values are analyzed in each results' subsection across defined intervals or ranges, enabling a more detailed comparison. 
By segmenting the data this way, the analysis highlights which hash function perform best within each operational range. 

%This approach assists designers in selecting hash functions based on specific resource or performance constraints relevant to their application domain.

\subsection{Execution Time Analysis}

\begin{figure}[tb]
    \centering
    \includegraphics[width=\textwidth]{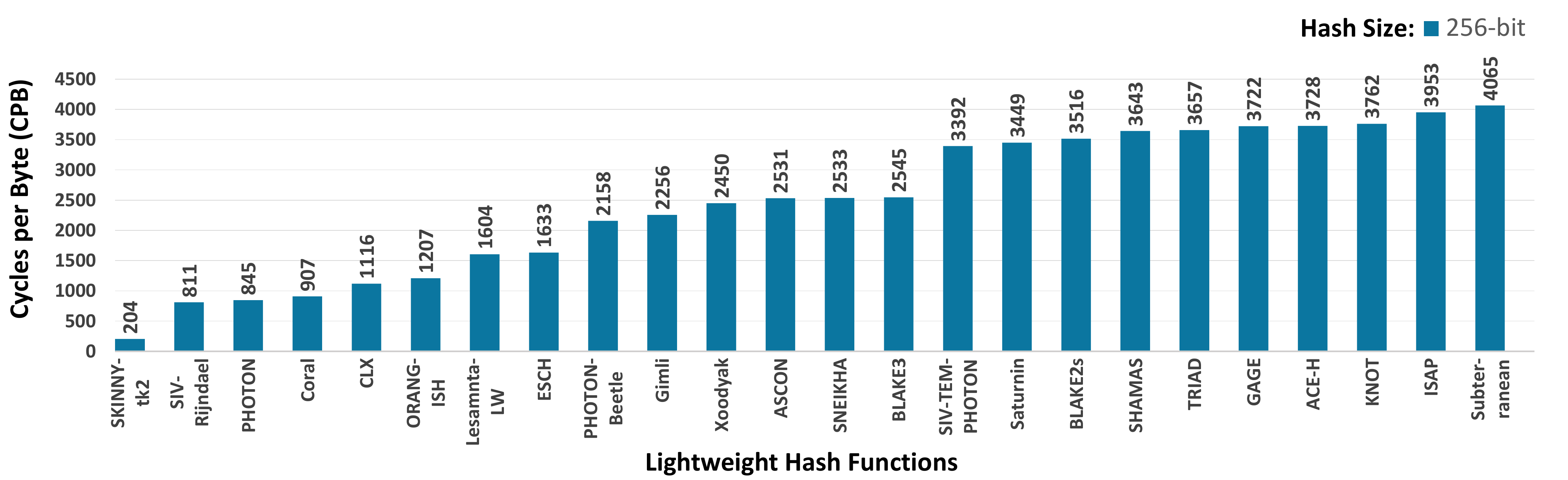}
    \caption{Execution time (\ac{CpB}) of lightweight hash functions.} \label{Fig:CpB}
\end{figure}

The execution time is quantified using the \ac{CpB} metric. 
A lower \ac{CpB} value indicates higher computational efficiency and faster throughput, whereas a higher \ac{CpB} corresponds to slower processing and lower throughput. 
Figure~\ref{Fig:CpB} illustrates that SKINNY-tk2 exhibits the lowest \ac{CpB} value, making it the most efficient hash function regarding execution speed. 
It is followed by SIV-Rijndael, which also demonstrates a highly optimized execution time. 
Within the $\leq$ 1000 \ac{CpB} range, the ISO-standardized PHOTON-256 also appears just after SIV-Rijndael with a slight margin. 
In the 1000--2000 \ac{CpB} range, the NIST-submitted candidates, including CLX and ORANGISH, appear first, while Lesamnta-LW, 
followed by another ISO-standardized hash function, Lesamnta-LW, with a significant margin. In the 2000--3000 \ac{CpB} segment, the execution time increases, with PHOTON-Beetle, Gimli, and Xoodyak preceding ASCON, followed by BLAKE3 with a slightly higher margin. 
In the 3000--4000 \ac{CpB} range, SIV-TEM-PHOTON appears first, followed by Saturnin, then the SHA-3 competition finalist BLAKE2s. Toward the upper limit, KNOT and ISAP exhibit higher \ac{CpB}. Finally, Subterranean shows the highest \ac{CpB}, 
indicating the lowest execution efficiency among the evaluated lightweight hash functions. 

\subsection{Memory Footprint Analysis}

\begin{figure}[tb]
    \centering
    \includegraphics[width=\textwidth]{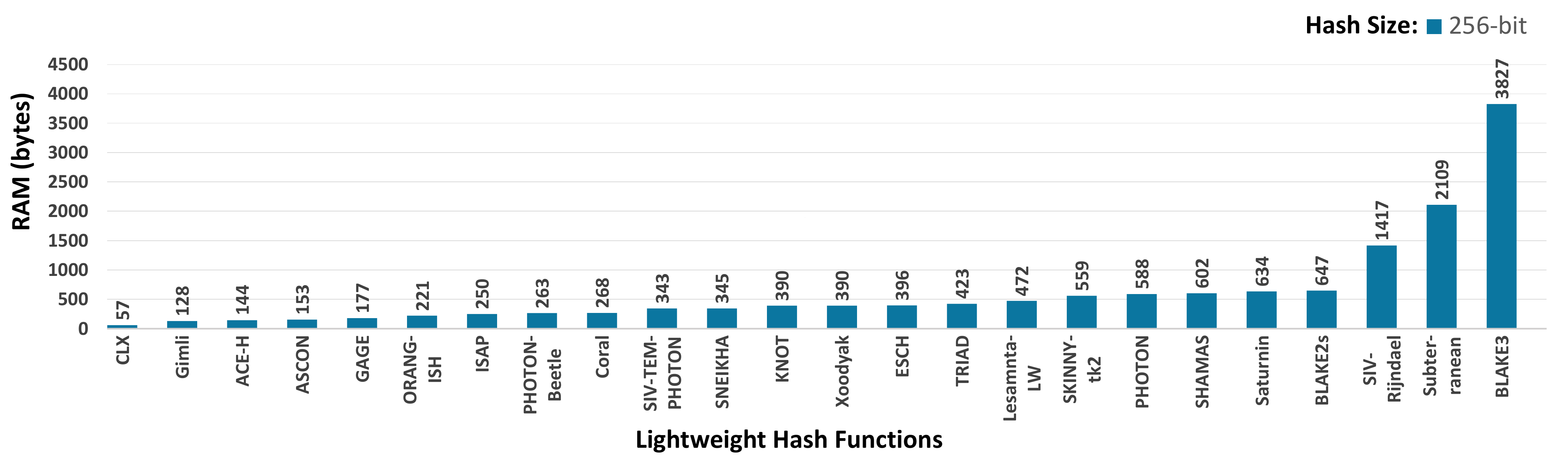}
    \caption{RAM usage of the lightweight hash function.} \label{Fig:RAM}
\end{figure}

\begin{figure}[tb]
    \centering
    \includegraphics[width=\textwidth]{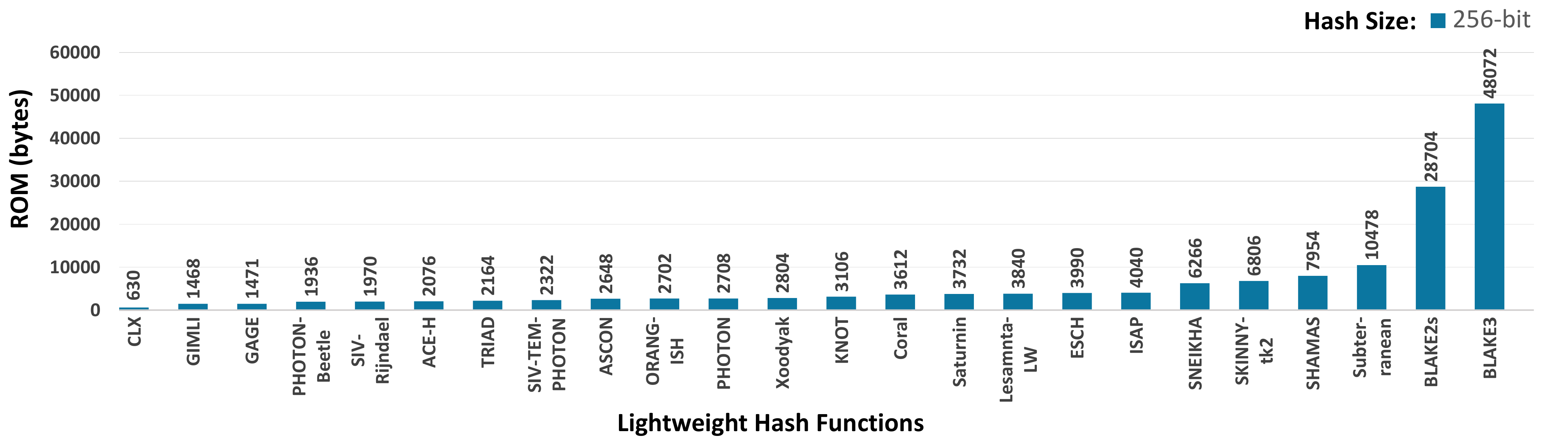}
    \caption{ROM usage of the lightweight hash functions.} \label{Fig:ROM}
\end{figure}

The memory footprint provides an estimate of the resource consumption of each hash function, evaluated in terms of RAM and ROM requirements.

\paragraph{RAM Consumption:} 
Figure~\ref{Fig:RAM} demonstrates that CLX, Gimli, and ACE-H exhibit the lowest RAM usage, 
making them highly suitable for devices with tight memory constraints. 
Within the $\leq 500$\,byte range, the ISO-standardized Lesamnta-LW has the highest RAM usage. 
In the 500--1000\,byte range, the NIST-submitted SKINNY-tk2 and the ISO-standardized PHOTON exhibit moderate RAM requirements, with BLAKE2s near the upper bound of this range and slightly higher than Saturnin. 
The most memory-intensive functions are BLAKE3, Subterranean, and SIV-Rijndael, which show significantly higher RAM consumption than the rest, with clear separation in their memory requirements.

\paragraph{ROM Consumption:} 
Figure~\ref{Fig:ROM} demonstrates that CLX and Gimli lead with the lowest ROM consumption as well. 
Within the $\leq 2000$\,byte range, SIV-Rijndael exhibits the highest ROM utilization, followed closely by PHOTON-Beetle. 
In the 2000--4000\,byte range, TRIAD and SIV-TEM-PHOTON consume the least, whereas ESCH and the Lesamnta-LW consume the most in this interval. 
PHOTON lies approximately at the midpoint of this range. Due to fewer candidates beyond this, the next interval spans 4000--10000\,byte, where ISAP and SNEIKHA show relatively modest ROM usage. 
In contrast, SHAMAS exhibits the highest usage in this range. Beyond \num{10000}\,byte, Subterranean has ROM consumption lower than both BLAKE2s and BLAKE3. 
Both variants of BLAKE are the most resource-intensive in terms of ROM.

These results demonstrate that CLX and Gimli offer highly optimized memory footprints, while BLAKE3 is the most memory-demanding in both RAM and ROM. 

% Designers targeting memory-constrained microcontrollers should prioritize candidates in the lower ranges, particularly those consistently efficient across both memory types.

\subsection{Energy Consumption Analysis}

\begin{figure}[tb]
    \centering
    \includegraphics[width=\textwidth]{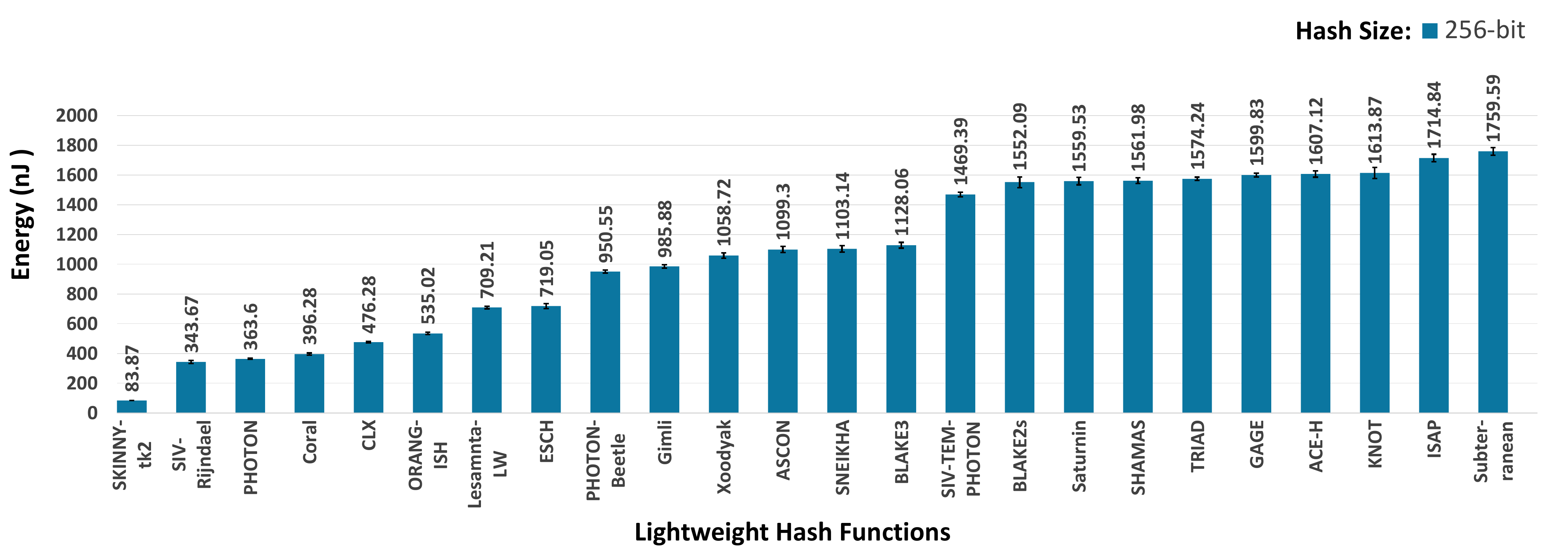}
    \caption{Energy consumption of selected lightweight hash function.} \label{Fig:Energy}
\end{figure}

Figure~\ref{Fig:Energy} plots the measured energy consumption of the evaluated hash functions, arranged in ascending order. 
Each data point is the mean of 10 runs; error bars indicate variability across executions.
The figure reflects the total energy required to process a fixed-length input and 
is particularly relevant for battery-powered or energy-constrained embedded systems.

The lowest energy consumption we observed is for SKINNY-tk2, followed by SIV-Rijndael.
While both are among the most energy-efficient, SKINNY-tk2 shows a notable margin of efficiency over SIV-Rijndael. 
In the $\leq 500$\,nJ range, CLX and Coral appear at the higher end, whereas PHOTON-256 demonstrates moderate energy consumption, occupying a middle position within this range. 
Within the 500--1000\,nJ range, ORANGISH and Lesamnta-LW exhibit lower energy usage, whereas Gimli and PHOTON-Beetle consume the most energy. 
In the 1000--1500\,nJ interval, Xoodyak and ASCON are positioned as the most efficient, whereas SIV-TEM-PHOTON consumes the highest energy in this range. 
For hash functions consuming more than 1500\,nJ, BLAKE2s and Saturnin are positioned at the lower end of this high-energy group. 
In contrast, Subterranean and ISAP exhibit the highest energy consumption across all evaluated candidates. 

\subsection{E-RANK Analysis}

\begin{figure}[tb]
    \centering
    \includegraphics[width=\textwidth]{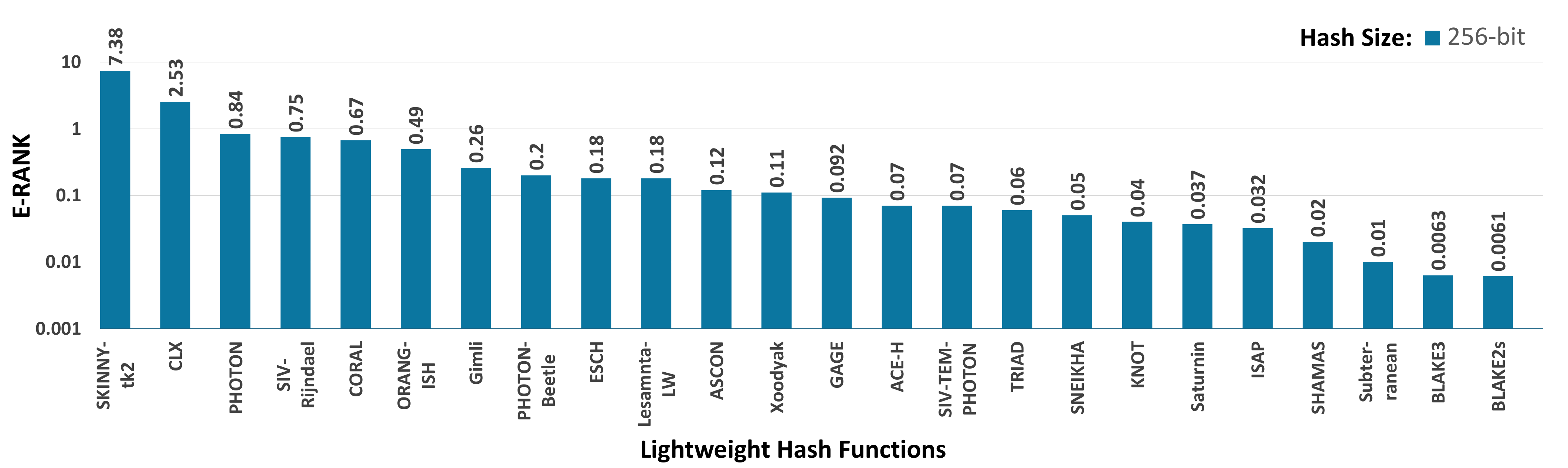}
    \caption{E-RANK of the studied lightweight hash functions.} \label{Fig:E-RANK}
\end{figure}

Figure~\ref{Fig:E-RANK} presents the $\modrank$ of the evaluated hash functions. The analysis reveals that SKINNY-tk2 achieves the highest $\modrank$ by a substantial margin. 
It is followed by CLX, which also attains an $\modrank$ greater than 1, demonstrating similarly strong efficiency characteristics. All remaining hash functions fall within the sub-unitary range ($\modrank$ < 1). 
Among these, PHOTON-256 exhibits the highest $\modrank$, closely followed by SIV-Rijndael, both reflecting favorable trade-offs under constrained-resource settings. 
Lesamnta-LW and ASCON achieve moderate $\modrank$ values, suggesting a reasonable balance, although not as optimized as the leading candidates. 
At the lower end of the spectrum, the BLAKE variants, specifically BLAKE2s, and BLAKE3, record the lowest $\modrank$ values among all evaluated hash functions.

%This outcome reflects their relatively high memory and energy demands and longer execution times on 8-bit architectures.

\subsection{Comparative Analysis}

Next, we present a comparative analysis of the benchmarked hash functions using a heatmap, which visually represents the efficiency of each metric.
For this, we first normalize memory, execution time, and energy consumption to a range between 0 and 1, using the min-max normalization method below, where 
the value $x$ is a single measurement and $X$ the set of measurements.

\begin{equation}
    \hat{x} = 1 - \frac{x - \min(X)}{\max(X) - \min(X)}
    \label{eq:inv_min_max_norm}
\end{equation}

\begin{figure}[tb]
    \centering
    \includegraphics[width=\textwidth]{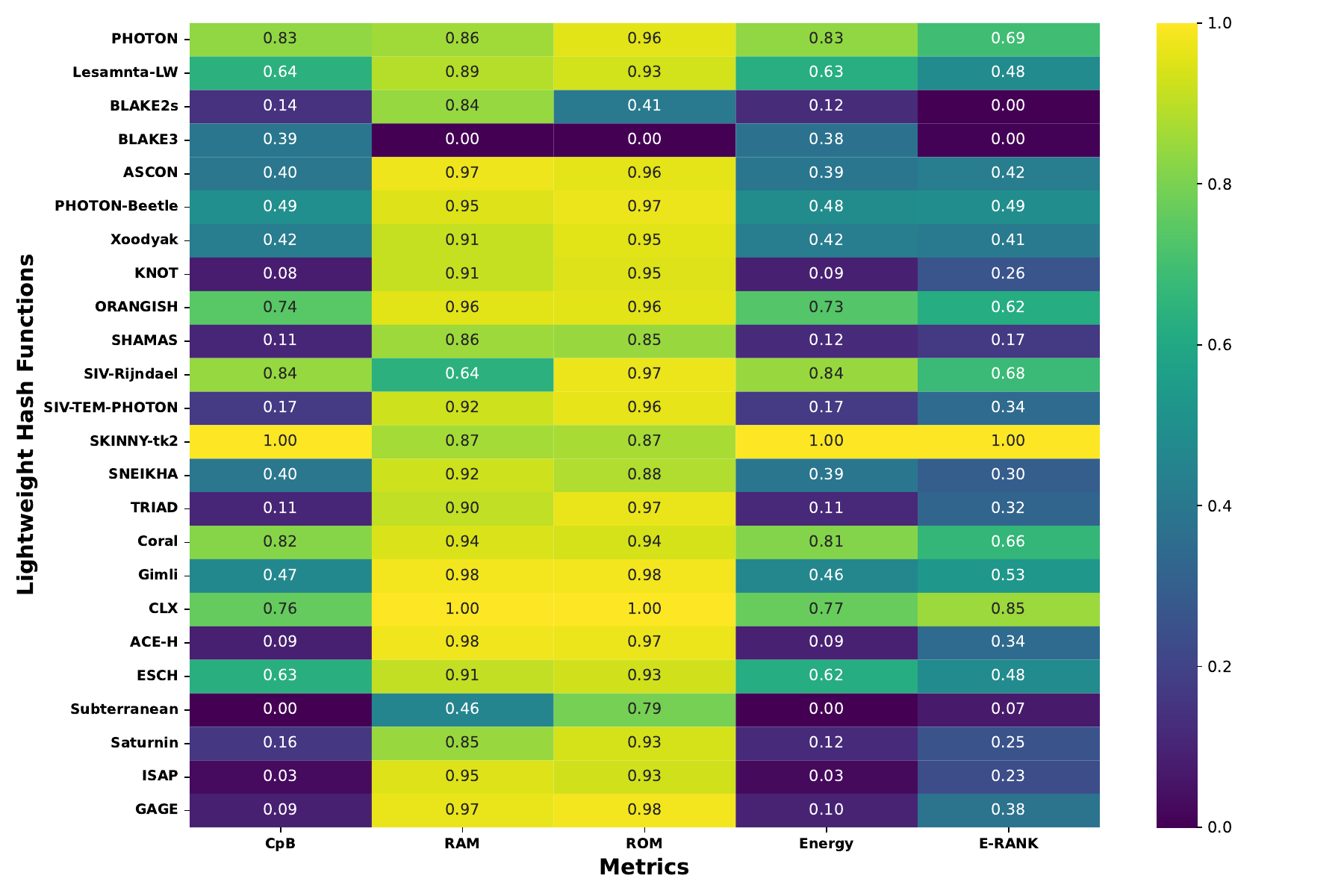}
    \caption{Normalized Heatmap of selected lightweight hash functions (Higher = Better).} \label{Fig:heatmap}
\end{figure}

For $\modrank$, values spans several orders of magnitude with many hash functions clustered closely together around the lower end of the scale.
To avoid having these values skewed and compressed by min-max normalization, making them hard to visualize in a heatmap,
we here apply a logarithmic transformation using the normalization formula below. The resulting heatmap is shown in Figure~\ref{Fig:heatmap}.
\begin{equation}
    \bar{x} = \frac{ \log\left(x \times  \min(X)\right)}{\log\left(\max(X)/\min(X)\right)}
    \label{eq:log_min_max_norm}
\end{equation}
%
%This transformation compresses extreme values, reduces skewness, and preserves the ranking order,
% making the visual representation clearer and more meaningful. 

In terms of \ac{CpB} and energy consumption, the evaluated hash functions SKINNY-tk2, SIV-Rijndael, PHOTON, and Coral exhibit the lowest \ac{CpB} and energy usage. 
\ac{CpB} is directly proportional to energy consumption. 
These hash functions utilize byte-aligned sponge constructions with fixed and non-branching permutations. 
They avoid runtime key scheduling, and the round functions consist of lightweight operations, such as 4-bit or 8-bit S-boxes, simple XORs, and linear diffusion layers. 
In contrast, ISAP, KNOT, and Subterranean have the lowest execution speeds and the highest energy consumption. 
This is primarily because these designs depend on bit-sliced or bit-level permutations, which consist of complex permutation structures. 
While these strategies enhance security and hardware optimization, they result in a higher \ac{CpB} in software, illustrating the trade-off between security and efficiency on constrained devices.

In terms of memory footprint, CLX and Gimli have the least memory usage. 
This is because these hash functions eliminate the need for dynamic memory allocation by utilizing fixed-size,
 byte-aligned sponge states, and their permutation cores rely on components such as XOR, rotations, 
 and Boolean logic instead of large S-boxes, lookup tables, or modular arithmetic.  
 In contrast, the hash functions with the highest memory footprints, including BLAKE3, BLAKE2s, and Subterranean 2.0, exhibit structural complexity. 
 BLAKE3 and BLAKE2s utilize complex compression schedules involving numerous constants and key-dependent initialization values, along with tree-based or parallel chunk processing and extendable output. Meanwhile, Subterranean employs bitwise transformations and supports multiple operational wrappers, all of which contribute to high memory requirements.

% Additionally, their implementations are static, unrolled, and free of branches, ensuring consistent control flow and compact code generation.

Some hash functions such as SKINNY-tk2, CLX, PHOTON, and SIV-Rijndael achieve high $\modrank$ scores due to their low \ac{CpB}, ROM, RAM, and energy consumption. 
Note that SKINNY-tk2, SIV-Rijndael, and PHOTON benefit specifically from low \ac{CpB} and energy usage. 
In contrast, CLX does not have the fastest execution speed but has the lowest memory footprint among these functions. 

Other hash functions, like BLAKE3, BLAKE2s, and Subterranean, show very low $\modrank$ scores due to their large memory footprints and high energy consumption. 
For example, the ROM and RAM requirements of BLAKE3 contribute to its lower $\modrank$, while Subterranean suffers from poor \ac{CpB} and energy efficiency combined with high RAM usage, which further reduces its $\modrank$. 
These findings emphasize that overall balanced performance is not determined only by execution speed, but rather by a balanced integration of processing rate, memory, and power metrics. 
This also suggests that faster execution generally necessitates more memory.

\begin{table}[t]
\centering
\scriptsize % Reduce font size to fit in one column
\renewcommand{\arraystretch}{1.2} % Adjust row height for better readability
\setlength{\tabcolsep}{5pt} % Reduce column spacing
\caption{Performance metrics calculated values of lightweight hash function.}
\begin{tabular}{l c c c c c}
\toprule
\textbf{LWHF} & 
\makecell{\textbf{CPB}} & 
\makecell{\textbf{RAM} \\ \textbf{(B)}} & 
\makecell{\textbf{ROM} \\ \textbf{(B)}} & 
\makecell{\textbf{Energy} \\ \textbf{(nJ)}} & 
\textbf{E-RANK} \\
\midrule

PHOTON & 845 & 588 & 2708 & 363.6 & 0.84\\
Lesamnta-LW & 1604 & 472 & 3840 & 709.21 & 0.18\\
BLAKE2s & 3516 & 647 & 28704 & 1552.09 & 0.0061\\
BLAKE3 & 2545 & 3827 & 48072 & 1128.06 & 0.0063\\
ASCON & 2531  & 153  & 2648  &  1099.3 & 0.12 \\ 
PHOTON-Beetle  & 2158   & 263  & 1936  & 950.55 & 0.2\\ 
Xoodyak & 2450  & 390  & 2804  & 1058.72 & 0.11\\ 
KNOT & 3762 & 390 & 3106 & 1613.87 & 0.04\\
ORANGISH & 1207 & 221 & 2702 & 535.02 & 0.49\\
SHAMAS & 3643 & 602 & 7954 & 1561.98 & 0.02 \\
SIV-Rijndael & 811 & 1417 & 1970 & 343.67 & 0.75\\
SIV-TEM-PHOTON & 3392 & 343 & 2322 & 1469.39 & 0.07\\
SKINNY-tk2 & 204 & 559 & 6806 & 83.87 & 7.38\\
SNEIKHA & 2533 & 345 & 6266 & 1103.14 & 0.05 \\
TRIAD & 3657 & 423 & 2164 & 1574.24 & 0.06 \\
Coral & 907 & 268 & 3612 & 396.28 & 0.67\\
Gimli & 2256 & 128 & 1468 & 985.88 & 0.26\\
CLX & 1116 & 57 & 630 & 476.28 & 2.53\\
ACE-H & 3728 & 144 & 2076 & 1607.12 & 0.07 \\
ESCH & 1633   & 396  & 3990  & 719.05 & 0.18\\ 
Subterranean & 4065 & 2109 & 10478 & 1759.59 & 0.01\\
Saturnin & 3449 & 634 & 3732 & 1559.53 & 0.037\\
ISAP & 3953  & 250  & 4040  & 1714.84 & 0.032\\ 
GAGE & 3722 & 177 & 1471 & 1599.83 & 0.092\\

\bottomrule
\end{tabular}
\label{tab:LWHF_performance}
\end{table}

%The research literature is particularly lacking comparative analysis of lightweight hash functions among themselves 
%or with other standardized \acp{LWHF} such as PHOTON, Lesamnta-LW, or BLAKE. 

\section{Related Work}

Few studies focus on benchmarking of software-based lightweight hash functions, other than the status reports published by NIST. 
There is a particular lack of comparative analysis with other standardized lightweight hash functions such as PHOTON, Lesamnta-LW, or BLAKE. 
To our knowledge, no other studies provide a quantitative evaluation or analytical assessment of the performance-to-cost tradeoff.

Khan et al.~\cite{Hw_benchmarking_NIST_lwhf} performed benchmarking emphasizing hardware implementations. 
The study evaluates the throughput-to-area (TP/A) ratio, hardware area utilization,
 and execution time but does not provide insights into the power analysis or software implementations.  
Windarta et al.~\cite{comparative_study} provide a detailed comparative study of NIST submitted hash functions, 
but the study concludes the results of multiple sources, such as previous research papers and internal evaluations by NIST. 
However, as a comparative study, it lacks a unified benchmarking approach to standardize performance comparisons.

The NIST status reports on lightweight cryptographic algorithms present software 
implementation results for NIST lightweight hash functions  in detail, 
evaluating execution time, memory footprint, and power consumption on ARM Cortex-M4, ESP32, 
and AVR ATmega328P~\cite{turan2023status,turan2021status,turan2019status}. 
However, these benchmarking results lack fine-grained power profiling, tradeoff analysis between performance and cost, 
and performance tuning for specific embedded platforms. 

\section{Conclusions}

This study presents a detailed methodology for conducting software benchmarking of lightweight hash functions 
on an AVR microcontroller using the ChipWhisperer platform. 
The experimental setup is designed to obtain precise measurements of key performance metrics. 
\ac{CpB} are accurately captured by reading the on-chip hardware cycle counter. 
The memory footprint is derived from post-compilation analysis using the AVR-GCC toolchain. 
Energy consumption is measured using a power measurement probe connected to the ChipWhisperer’s integrated oscilloscope.

A comprehensive comparison between ISO-standardized lightweight hash functions, such as PHOTON and Lesamnta-LW,
 and NIST-submitted lightweight hash functions, including SHA competition finalists from the BLAKE family, reveals several trends. 
 The hash functions from the ISO and NIST-LWC portfolios show that faster execution generally requires more memory (e.g., SKINNY-tk2), 
 while designs that focus on security tend to execute more slowly (e.g., ISAP). 
 ISO-standardized designs like PHOTON and Lesamnta-LW were initially created with hardware implementation in mind, 
 but the results show that these hash functions are well-suited for software implementation on 8-bit platforms. 
 Also, many NIST-LWC submissions, such as SKINNY, CLX, and SIV-Rijndael, strike a strong balance across performance metrics. 
 The BLAKE2s and BLAKE3 variants of the BLAKE family have a strong structure for security but tend 
 to perform poorly on AVR-constrained platforms due to their high memory requirements and considerable amount of execution speed and energy consumption.
 Thus, there is currently no single optimal lightweight hash function for all scenarios; 
 the best choice depends on specific application requirements and the architecture of different microcontrollers for real-world deployments.

%\textbf{\ackname}{This research received no external funding.}

%\textbf{\discintname}{The authors have no conflicts of interest to declare.}

%\printbibliography

\bibliographystyle{splncs04}
\bibliography{main.bbl}

\end{document}